\documentclass[%
 reprint,
 amsmath,amssymb,
 aps,
pra,
]{revtex4-1}
\usepackage{float}
\usepackage{graphicx}
\usepackage{dcolumn}
\usepackage{bm}

\usepackage{xcolor}
\usepackage[normalem]{ulem}
\newcommand{\ave}[1]{\left \langle #1 \right \rangle}
\newcommand{\ket}[1]{\left | #1 \right \rangle}

\newcommand{\mele}[3]{\left \langle #1 \middle | #2 \middle | #3 \right \rangle}

\definecolor{commentColorMP}{rgb}{0.0,0.2,0.6}
\definecolor{commentColorZW}{rgb}{0.6,0.2,0.0}
\definecolor{commentColorASN}{rgb}{0.2,0.6,0.0}

\newcommand{\omegap}{{\omega_{\text{p}}}}
\newcommand{\omegac}{{\omega_{\text{c}}}}
\newcommand{\omegad}{{\omega_{\text{d}}}}
\newcommand{\phip}{{\phi_{\text{p}}}}
\newcommand{\phid}{{\phi_{\text{d}}}}
\newcommand{\omegacmax}{{\omega_{\text{c,max}}}}
\newcommand{\GHz}{{\text{GHz}}}
\newcommand{\td}{t_{\text{d}}}

\newcommand{\opa}{\hat{a}}
\newcommand{\opad}{\hat{a}^\dagger}
\newcommand{\EC}{E_{\text{C}}}
\newcommand{\EJ}{E_{\text{J}}}
\newcommand{\EJmax}{E_{\text{J,max}}}

\begin{document}

\preprint{APS/123-QED}

\title{Quantum dynamics of a few-photon parametric oscillator}

\author{Zhaoyou Wang}
\thanks{These two authors contributed equally}
\author{Marek Pechal}%
\thanks{These two authors contributed equally}
\author{E. Alex Wollack}%
\author{Patricio Arrangoiz-Arriola}%
\author{Maodong Gao}%
\author{Nathan R. Lee}%
\author{Amir H. Safavi-Naeini}%
 \email{safavi@stanford.edu}
\affiliation{%
 Department of Applied Physics and Ginzton Laboratory, Stanford University\\
 348 Via Pueblo Mall, Stanford, California 94305, USA
}%

\date{\today}

\begin{abstract}
Modulating the frequency of a harmonic oscillator at nearly twice its natural frequency leads to amplification and self-oscillation. Above the oscillation threshold, the field settles into a coherent oscillating state with a well-defined phase of either $0$ or $\pi$. We demonstrate a quantum parametric oscillator operating at microwave frequencies and drive it into oscillating states containing only a few photons. The small number of photons present in the system and the coherent nature of the nonlinearity prevents the environment from learning the randomly chosen phase of the oscillator. This allows the system to oscillate briefly in a quantum superposition of both phases at once --  effectively generating a nonclassical Schr\"{o}dinger's cat state. We characterize the dynamics and states of the system by analyzing the output field emitted by the oscillator and implementing quantum state tomography suited for nonlinear resonators. By demonstrating a quantum parametric oscillator and the requisite techniques for characterizing its quantum state, we set the groundwork for new schemes of quantum and classical information processing and extend the reach of these ubiquitous devices deep into the quantum regime.
\end{abstract}

\pacs{Valid PACS appear here}
\maketitle


Parametric amplifiers and oscillators are quintessential devices used to amplify small electromagnetic signals \cite{Castellanos-Beltran2008,Macklin2015,Krantz2016}, convert radiation from one frequency to another \cite{Harris1969,Myers1995},  create squeezed light and entangled photons \cite{Yurke1988,Schnabel2017}, and realize new information processing architectures \cite{Mcmahon2016, Puri2017, Puri2018, Goto2016, Goto2018}. They operate by modulating a parameter, the natural frequency of the resonant circuit $\omegac$, at approximately twice its frequency $\omegap\approx 2\omegac$. The modulation amplifies one of the field quadratures at the half-harmonic frequency $\omegap/2$. A sufficiently large amplification overtakes the detuning and decay present in the system and leads to an exponential increase in the half-harmonic cavity field amplitude. Nonlinearities clamp this exponential growth and cause the system to enter an oscillating steady-state. These nonlinearities can be dissipative or dispersive. An example of the latter is the Kerr nonlinearity that induces a change in the cavity frequency proportional to the intracavity field intensity or photon number. Furthermore, the self-oscillation amplitude scales inversely with the magnitude of the nonlinearity.  These nonlinearities have been exceedingly small in parametric oscillators to date, leading to large oscillation amplitudes that result in rapid decay of quantum coherence and the appearance of classical dynamics~\cite{Wilson2010}. The quantum regime of nonlinear parametric oscillators has been extensively studied in theory \cite{Kinsler1991,Wustmann2013,Zhang2017} but received only limited attention in experiments \cite{Ding2017}. 

\begin{figure}[h]
    \centering
    \includegraphics[width=0.4\textwidth]{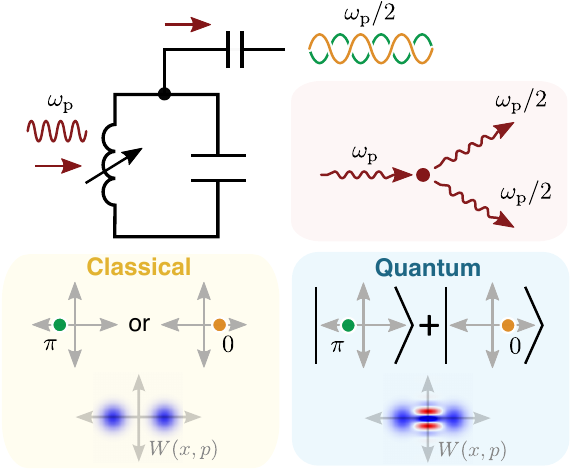}
    \caption{{\bf Schematic of a parametric oscillator above the instability threshold.} An $LC$ circuit with a harmonically modulated inductance is an example of a parametric oscillator. Above threshold, the system is oscillating at half the driving frequency $\omega_{\text{p}}/2$, which can be described as a parametric down-conversion process. The phase of the oscillator with respect to the external driving can be $0$ or $\pi$. Classically, this symmetry is broken and one of the two cases is realized at random. Quantum mechanically, the oscillator can be in a superposition of both states.}
\end{figure}

\begin{figure*}[t]
   \includegraphics[width=0.7\textwidth]{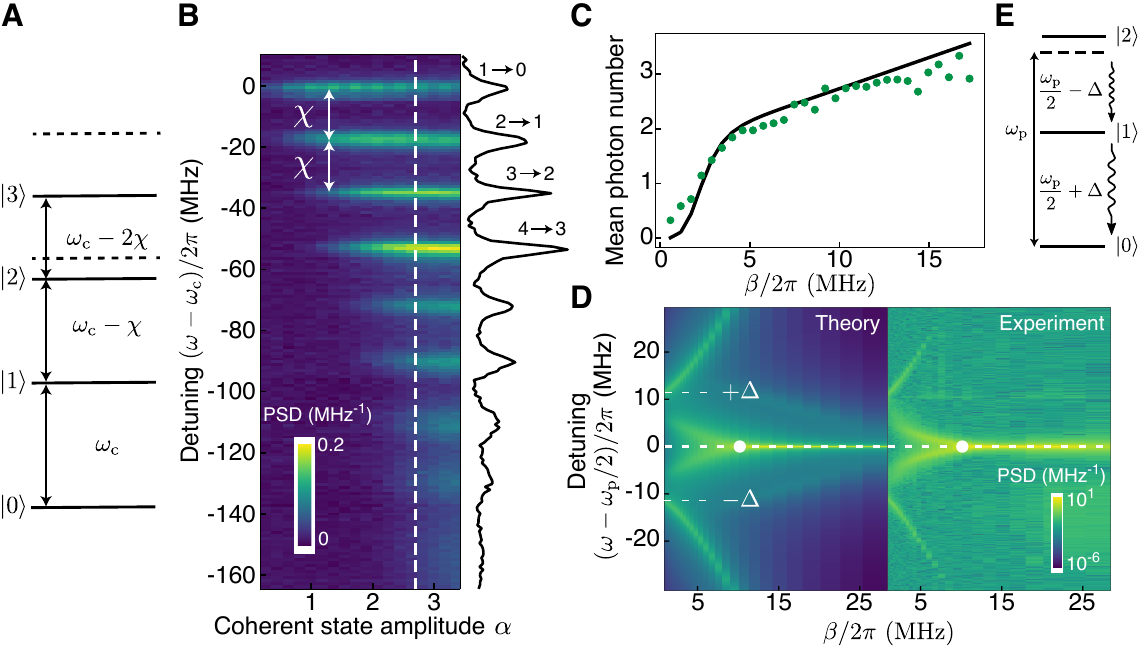}
   \caption{{\bf Nonlinear energy level structure and steady state dynamics of KPO.} ({\bf A}) Energy level diagram for Kerr nonlinear resonator with each additional photon reducing the transition frequency by $\chi$. ({\bf B}) Transient PSD measurements for coherent states with different amplitudes $\alpha$. The curve plotted on the right is the PSD corresponding to $\alpha$ indicated by the white dashed line. ({\bf C}) Mean photon number $\ave{\hat{n}}$ at steady state versus the amplitude of the parametric driving at $\Delta/2\pi = 25.3$ MHz. The quantum prediction (solid line) agrees well with the experimental data. ({\bf D}) Steady state PSD (logarithmic scale) for different driving amplitudes at $\Delta/2\pi = 11.2$ MHz. The spectrum goes from a double-peaked shape at small $\beta$  to a single narrow peak in the self-oscillation regime, passing through a multi-peaked spectrum at intermediate values of $\beta$. The white dot indicates the approximate on-set of self-oscillation. The spacing between the two peaks in the weak drive limit is $2\Delta$, as explained by the two-photon emission process shown in ({\bf E}).}
\end{figure*}

We experimentally realize a quantum Kerr parametric oscillator (KPO) by implementing an on-chip  superconducting nonlinear resonator and investigate its quantum dynamics under parametric driving. In contrast to previously demonstrated optical and microwave parametric oscillators, our device operates in the quantum regime with a self-oscillating state containing only a few photons. The resonator is implemented as an $LC$ circuit with an array of Josephson junctions in place of the inductor \cite{Castellanos-Beltran2007}. The nonlinear inductance of this array induces a Kerr interaction $-\frac{\chi}{2}\opad\opad\opa\opa$, where $\opa$ is the annihilation operator of the resonator and $\chi/2\pi = 17.3\,\mathrm{MHz}$ is the resonator frequency shift per photon (Fig.~2A). The linewidth of the resonator is  $\kappa/2\pi \approx 1.1\,\mathrm{MHz}$, which means that we are well within the single-photon Kerr regime \cite{Kirchmair2013} with $\chi/\kappa \approx 17$. The resonator frequency $\omega_{\text{c}}/2\pi$ can be tuned down from $8~\GHz$ to below $4~\GHz$ by an on-chip flux line and most of the measurements are done with the resonator in the $6-8~\GHz$ frequency range. This tunability also enables parametric driving of the form $\tilde{\beta}(t)(\opa+\opad)^2$ where $\tilde{\beta} (t)$ is proportional to the voltage $V(t)$ applied to the flux line.

With parametric driving $\tilde{\beta} (t) = 2\beta(t) \cos \omega_{\text{p}}t$ at frequency $\omega_{\text{p}} = 2(\omega_{\text{c}}-\Delta)$, which is slightly detuned from the parametric resonance $2\omega_{\text{c}}$, the dynamics of the resonator in a rotating frame at half the driving frequency is well described by the Hamiltonian
\begin{equation}
    \hat{H}/\hbar = \Delta \opad\opa - \frac{\chi}{2} \opad\opad\opa\opa + \beta(t) (\opa^2 + \opa^{\dagger 2})
\end{equation}
where $\beta(t)$ is the slowly varying amplitude of the parametric driving. 

We first characterize the energy level structure of the Kerr nonlinear resonator without parametric driving ($\beta=0$). To do this, we prepare the resonator in an excited state $\hat{\rho}_0$ and let it relax back to the vacuum state while we collect the emitted microwave signal. The power spectral density (PSD) of this signal, which we call ``transient PSD" to distinguish it from the PSD measured at steady state, contains multiple peaks evenly spaced by $\chi$ (Fig.~2B) due to the nonlinear energy level structure. The $n$th peak captures photons at frequency $\omega_c-(n-1)\chi$, emitted during relaxation from $\ket{n}$ to $\ket{n-1}$. We measure the transient PSDs for coherent states $\hat{\rho}_0 = |\alpha\rangle\langle\alpha|$, which we prepare by driving the system resonantly at $\omega_c$ with 1~ns pulses of different amplitudes. Since the pulses are much shorter than $1/\chi$, they simply displace the state of the resonator from vacuum into a coherent state  $\ket{\alpha}$, with $\alpha$ proportional to the pulse amplitude. In agreement with theory, the number of peaks we observe in the spectrum grows with $\alpha$ as higher Fock states are populated (Fig.~2B).

The transient PSD measurement provides a way to infer the Fock occupations $ p_n = \mele{n}{\hat{\rho}_0}{n}$ of the initial state $\hat{\rho}_0$. The probability that a transition from $\ket{n}$ to $\ket{n-1}$ occurs during free relaxation to $|0\rangle$ is exactly equal to the total population of all Fock states $\ket{n}$ and higher. When the peaks are resolved $(\chi\gg\kappa)$, the total power in the $n$th peak is proportional to $\sum_{k=n}^{\infty}p_{k}$ (see SI). The measured signal is related to the field emitted from the resonator by a frequency-dependent gain factor. The calibrated gain allows us to calculate the Fock occupations $\{ p_n \}$ in the initial state from the transient PSD. 

The quantum parametric oscillator does not have a sharp self-oscillation threshold. To study the transition to self-oscillation, we modulate the flux line with a continuous wave at a fixed frequency $\omega_{\text{p}} = 2(\omega_{\text{c}}-\Delta)$ and measure the mean photon number (Fig.~2C) as well as the steady state PSD (Fig.~2D) for increasing parametric drive amplitudes $\beta$. 
The mean photon number $\langle \hat{n}\rangle$ (Fig.~2C) is obtained by turning the drive off and measuring the transient PSDs as the state relaxes. At large $\beta$, both the classical and the quantum model predict photon numbers very close to the measured results. For smaller $\beta$, the measurements show a gradual transition into self-oscillation, smoothed by quantum fluctuations. This behavior deviates from the classical prediction but is reproduced well by the quantum model. 

The steady state PSD (Fig.~2D) spectrum reduces to two peaks spaced by $2\Delta$ in the weak drive limit and can be understood as the result of a two-photon emission process, with one photon emitted at the cavity frequency $\omega_{\text{c}} = \omega_{\text{p}}/2 + \Delta$ and the other, by energy conservation, at $\omega_{\text{p}}/2 - \Delta$ (Fig.~2E).  For large drive amplitude $\beta$ the resonator enters the self-oscillation regime and is phase-locked to the external parametric drive leading to the narrow peak seen in Fig.~2D. The narrowing of this peak results from the reduction of the switching rate between the two possible self-oscillating states ($0$ and $\pi$ phase) as we increase $\beta$. 

\begin{figure}[h]
    \includegraphics[width=0.47\textwidth]{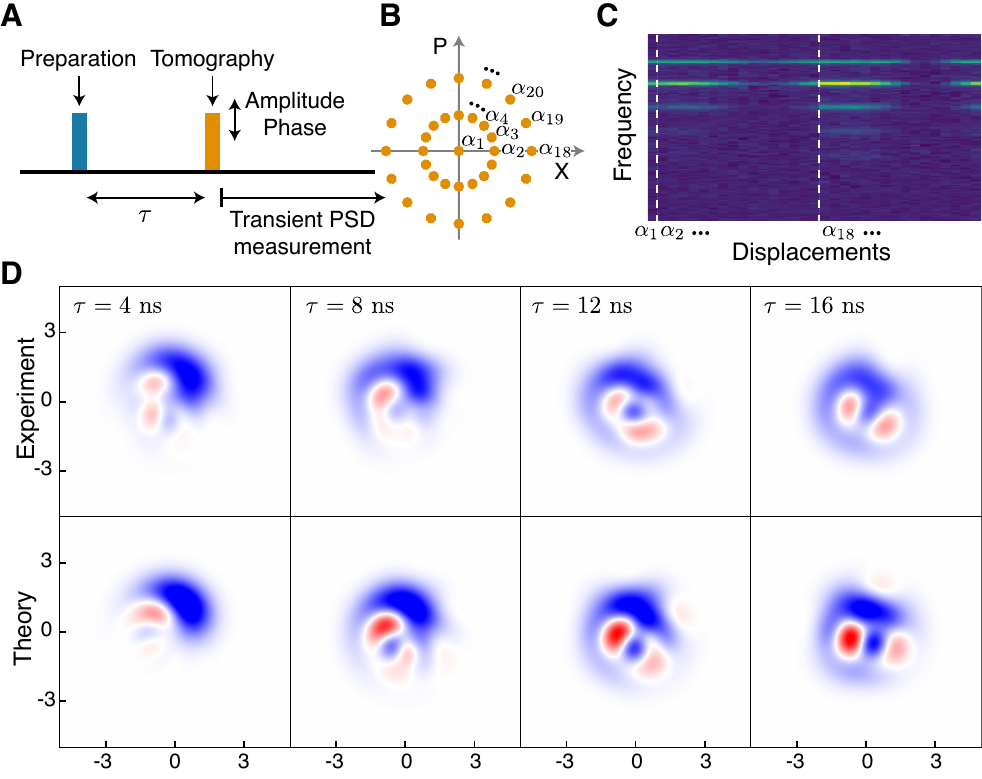}
    \caption{{\bf Schematic of the tomography and time evolution of a coherent state under Kerr nonlinearity.} ({\bf A}) The tomography pulses are delayed by $\tau$ from the state preparation pulse and ({\bf B}) swept in phase space with 2 different amplitudes, which corresponds to displacements with 16 different phases and two different amplitudes of about $0.5$ and $1.0$. ({\bf C}) An example tomography dataset for a coherent state ($\tau=0$). ({\bf D}) Time evolution of a coherent state at four different time slices. The distortion in phase space is caused by the Kerr nonlinearity and is reproduced by simulations. The fidelities of the reconstructed states compared with simulated ones are 0.93, 0.92, 0.90, 0.89, respectively.}
\end{figure}

Characterizing the full quantum state of the parametric oscillator is challenging and calls for a distinct approach to quantum tomography. There are several methods based on measuring Fock state {populations} or parities of displaced states. These approaches were developed for qubit-resonator systems~\cite{Lutterbach1997,Hofheinz2009,Kirchmair2013,Vlastakis2013} or qubits with tunable coupling to the environment~\cite{Shalibo2013} and require more complicated devices with auxiliary cavities and control parameters. Other methods have been developed in quantum optics based on studying the statistics of the output field $\hat{a}_{\text{out}}$ of linear resonators \cite{Lvovsky2001,DaSilva2010,Eichler2011}, where $\hat{a}_{\text{out}}$ is {linearly} related to the resonator field $\hat{a}(t=0)$ at some initial time. These methods are not suitable for our system since $\chi\gg\kappa$ leads to a highly nonlinear relationship between the output field and the intracavity mode. 


We develop a state tomography method suited for the quantum parametric oscillator that does not need any auxiliary systems. Our method is based on measuring the transient PSD using a purely linear detection of the output field. The transient PSD measurement provides us with the diagonal elements of $\hat{\rho}$. By displacing the state in phase space and detecting the transient PSD, we find the diagonal elements of a displaced density matrix $\hat{D}\hat{\rho}\hat{D}^\dagger$ that contain information about the off-diagonal elements of $\hat{\rho}$. Repeating this for several different displacements, we obtain enough information to estimate the full density matrix by a maximum likelihood method. This aspect of the tomography method is conceptually similar to an existing technique using generalized $Q$ functions \cite{Shen2016,Kirchmair2013}. 
We arrive at the estimate of $\hat{\rho}$ by minimizing a loss function $\mathcal{L}(\hat{\rho}_\text{est})$, which quantifies the difference between the measured PSD and the PSD simulated for the state $\hat{\rho}_\text{est}$ (see SI).
This convex optimization problem can be solved efficiently by semi-definite programming. The broadening of the higher Fock state peaks due to their shorter lifetime limits the maximum size of an unknown state that we can reconstruct. Additionally, we have observed systematic errors which become more severe for larger measured states. We suspect this to be due to excitation of unwanted off-resonant transitions by strong drive pulses and other nonlinear effects.

\begin{figure*}[t]
    \centering
    \includegraphics{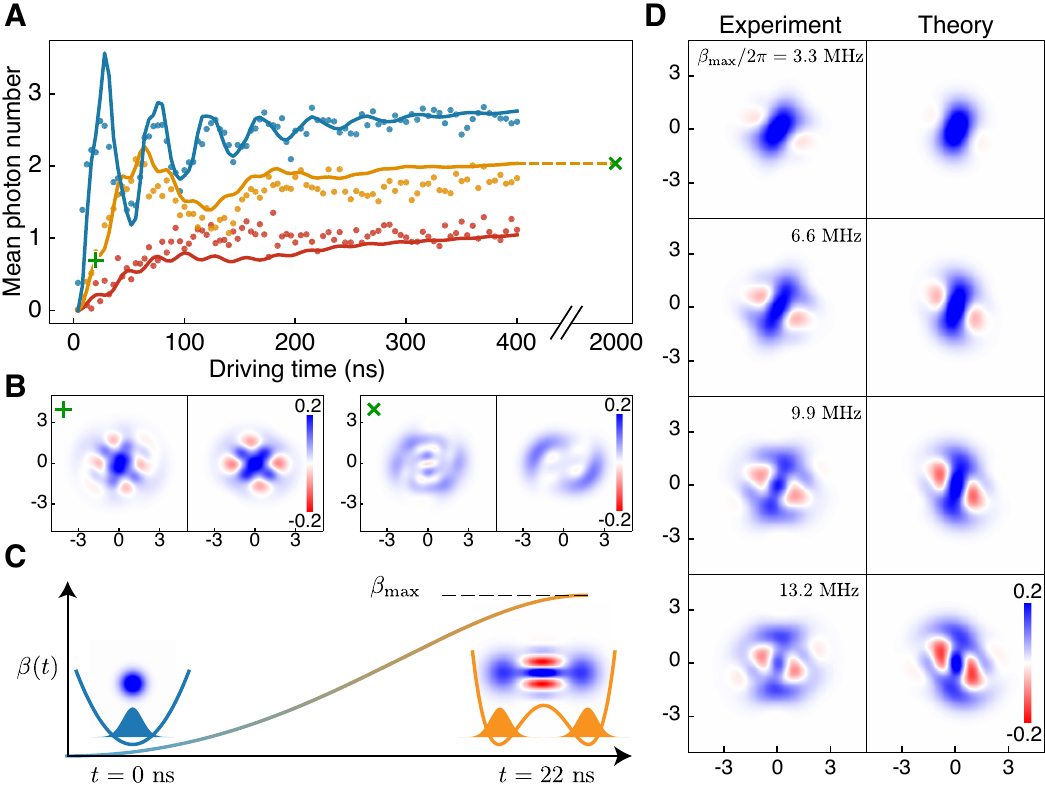}
    \caption{{\bf Transient state dynamics of the parametrically driven KPO.} ({\bf A}) Time evolution of the mean photon number up to 400 ns at $\Delta/2\pi = 24.6\,\text{MHz}$ for $\beta/2\pi=3.5\,\text{MHz}$ (red), $\beta/2\pi=5.8\,\text{MHz}$ (yellow), $\beta/2\pi=11.5\,\text{MHz}$ (blue). ({\bf B}) Both state reconstructions at $t=20\,\text{ns}$ (transient state) and $t=2000\,\text{ns}$ (steady state) show good agreement with theory (left: experiment, right: theory) with fidelities of $0.93$ and $0.94$ respectively. ({\bf C}) Pulse amplitude profile $\beta (t)$ for cat state generation. A pulse length of $22\,\text{ns}$ is chosen from simulation. Ramping $\beta$ up from $0$ to $\beta_{\text{max}}$ drives the resonator from the vacuum state into an oscillating cat state. ({\bf D}) Tomography results of generated cat states for different $\beta_{\text{max}}$ at $\Delta/2\pi = -6.7\,\text{MHz}$, which match closely (fidelities $0.98$, $0.95$, $0.92$, $0.89$) with simulations.}
\end{figure*}

We use this tomography procedure to study the free dynamics of the system. After displacing the resonator from vacuum to a coherent state, we let the system evolve freely for a time $\tau$ before performing the tomography (Fig.~3A). Here as well as in the rest of the paper, the displacements used in the tomography procedure are arranged in concentric rings as illustrated in Fig.~3B. Fig.~3C shows an example of a raw tomography dataset, consisting of the transient PSDs for all displacements. We reconstruct the state at different time slices $\tau$ and observe the evolution of the Wigner distributions due to the Kerr nonlinearity (Fig.~3D). In agreement with simulations, this evolution involves a ``shearing" distortion, which has a classical counterpart explained by the amplitude-dependence of the resonator frequency, and also exhibits negative Wigner function values, a signature of quantum mechanical behavior~\cite{Kirchmair2013}.

The system can behave nonclassically under a parametric drive, but only before photons leaking out of the oscillator cause the loss of quantum coherence. In our system, quantum dynamics persist in this transient regime for a sufficiently long time to allow  detailed observation. This nonclassical evolution is important to understand in the context of emerging applications for parametric oscillators in quantum information processing~\cite{Goto2016,Puri2018}. 
To investigate the transient dynamics, we turn on the parametric drive suddenly and measure the time evolution of the mean photon number for three different drive amplitudes, $\beta/2\pi = 3.5, 5.8$ and $11.5~\text{MHz}$ (Fig.~4A). The observed time dependence of the mean photon number is in good agreement with a theory fit to all three data sets simultaneously, using only the detuning $\Delta$, the loss rate $\kappa$ and a drive conversion factor $\beta/V$ as free fit parameters. Using the previously described tomography procedure, we also reconstruct the quantum state for $\beta/2\pi=5.8\,\text{MHz}$ at $t = 20\,\text{ns}$ (transient state) and $t=2000\,\text{ns}$ (steady state) (Fig.~4A). Comparison between the theoretically predicted and experimentally obtained Wigner functions shows relatively good agreement with fidelities of $0.93$ and $0.94$ for the transient and steady states respectively. We attribute the discrepancies to systematic errors in the tomography process which we believe could be mitigated by improving its calibration. The only fit parameters in the theoretical prediction are the overall rotation angle and a short delay ($t_\text{d}=2.5\,\text{ns}$) between the end of the parametric drive pulse and the start of the tomography (see SI).

These results demonstrate that the state of the oscillator can be engineered by designing the parametric drive $\beta(t)$. For example, by adiabatically changing $\beta(t)$, we can prepare even-parity energy eigenstates of the Hamiltonian for different values of $\beta$ as long as losses are negligible \cite{Zhang2017}. Intriguingly, for $\Delta<0$, as $\beta$ approaches and exceeds $\chi$, the energy eigenstate adiabatically connected to the vacuum state begins to closely approximate the even-parity Schr\"{o}dinger's cat state. We set the pump detuning in the experiment to $\Delta/2\pi=-6.7~\text{MHz}$, and begin with the resonator in the vacuum state with $\beta=0$. We increase $\beta$ slowly so the system follows the eigenstate of the Hamiltonian.
Numerical simulations suggest that given our system's parameters,
\begin{equation}
    \beta (t) = \beta_{\text{max}} \sin^2 \frac{\pi t}{2 t_{\text{max}}}, \quad 0\leq t \leq t_{\text{max}},
\end{equation}
with $t_{\text{max}}=22~\text{ns}$ (Fig.~4B) is a reasonable parametric driving profile for preparing a cat state. The length of this signal is much shorter than the cavity decay time $1/\kappa \approx 150\,\mathrm{ns}$ but long enough to ensure approximately adiabatic evolution. We perform the experiment ramping to different values of $\beta_{\text{max}}$ and verify the result with the state tomography procedure. The results are compared to the Wigner functions found theoretically (Fig.~4D). In the simulations, the rotation angle and the drive conversion factor $\beta/V$ are the only fit parameters and common to all four data sets. As described above, we again assume a short ($t_\text{d}=2.5~\text{ns}$) period of free evolution between the end of state preparation and the start of our tomography, which causes a small distortion (due to $\chi$) of the reconstructed state with respect to the eigenstate of the driven system. In our experiment, we ramp the drive up to $\beta_{\text{max}}\sim\chi$, which is necessary to see the emergence of the cat state. To a good approximation, the generated states after the short free evolution, described by $\hat{U}_0(t_\text{d})$, are 
\begin{equation}
\ket{\psi(\alpha)}\propto\hat{U}_{0}(t_\text{d})(\ket{\alpha}+\ket{-\alpha}),
\end{equation}
with $\alpha=0.64, 0.88, 1.08, $ and $1.2$ for the data shown in Fig.~4D. The largest of these corresponds to a $4|\alpha|^2 = 5.8$ photon Schr\"odinger's cat state~\cite{Deleglise2008,Vlastakis2013}. In the $\beta_{\text{max}}\gg\chi$ regime, the relevant eigenstate exponentially approaches a cat state of size $8\beta/\chi$ due to the double-well shape of the system's effective potential (Fig.~4C) (see SI for more details). For very large cat states, imperfections in the tomography process that grow with the number of photons in the analyzed state prevent its faithful reconstruction. In comparison to other schemes for cat state generation~\cite{Deleglise2008,Leghtas2013,Vlastakis2013,Leghtas2015,Davis2018}, our method is significantly more hardware efficient as it requires only a resonator with one input and one output line for both state generation and read-out.

The parametric oscillator is one of the paradigmatic systems in quantum optics and has found an enormous range of applications over the years. We have experimentally demonstrated the few-photon quantum dynamics of a parametric oscillator by introducing a large Kerr nonlinearity in the microwave frequency regime. We show that the system can adiabatically generate cat states of five to six photons and have developed a tomography method suited for the  characterization of its state and dynamics. Our work demonstrates that nontrivial quantum states can be engineered and characterized with nearly minimal hardware complexity. The quantum coherence and hardware efficiency of the system bode well for the prospects of scaling these devices to larger networks in emerging applications of parametric oscillators for quantum information processing and optimization~\cite{Mcmahon2016, Puri2017, Puri2018, Goto2016, Goto2018}.

\section*{Acknowledgments}

This work was supported by the US Department of Energy through grant number DE-SC0019174. A.S.-N.  acknowledges support from the David and Lucille Packard Fellowship. E.A.W. was supported by a National Defense Science and Engineering Graduate (NDSEG) Fellowship. P.A.A. and N.R.L. were supported by Stanford Graduate Fellowships (SGF). M.G. acknowledges support from the Tsinghua University undergraduate research program. M.P. acknowledges support from the Swiss National Science Foundation. Part of this work was performed at the Stanford Nano Shared Facilities (SNSF), supported by the National Science Foundation under Grant No. ECCS-1542152, and the Stanford Nanofabrication Facility (SNF).



\pagebreak
\begin{center}
	\textbf{\large Supplemental Materials: Quantum dynamics of a few-photon parametric oscillator}
\end{center}
\setcounter{equation}{0}
\setcounter{figure}{0}
\setcounter{table}{0}
\makeatletter
\renewcommand{\theequation}{S\arabic{equation}}
\renewcommand{\thefigure}{S\arabic{figure}}
\renewcommand{\bibnumfmt}[1]{[S#1]}
\renewcommand{\citenumfont}[1]{S#1}

\section{Materials and methods}
\subsection{Device fabrication}
\begin{figure}
	\centering
	\includegraphics[scale = 0.4]{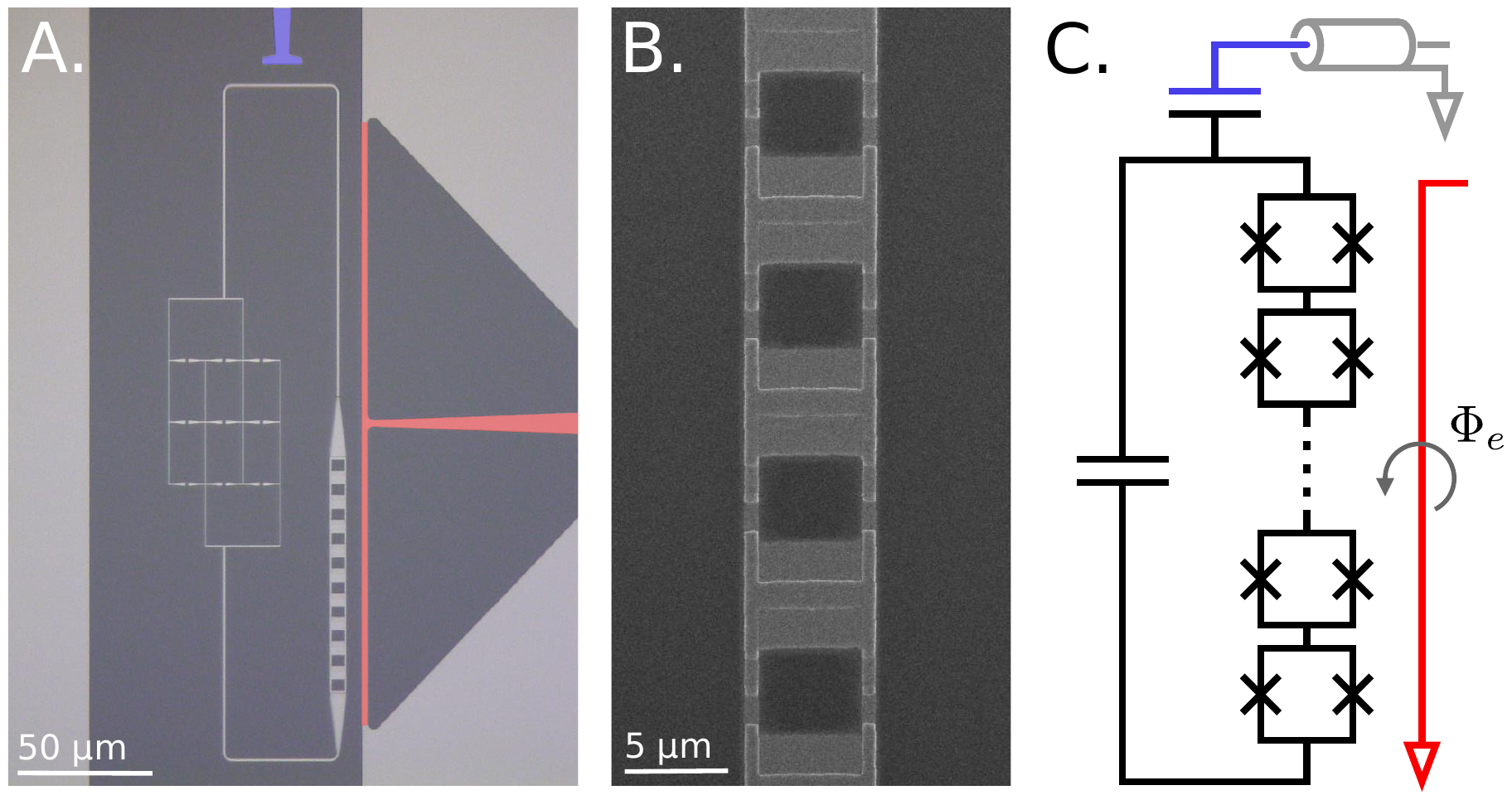}
	\caption{ {\bf Device fabrication.} ({\bf A}) False colored micrograph of the Kerr parametric oscillator. The coupling capacitor and flux bias line are shown in blue and red, respectively. The capacitor shunting the SQUID array is shaped in an unconventional way in order to accommodate an array of nanomechanical resonators introduced in later devices \cite{arrangoiz18}. ({\bf B}) SEM image of the SQUID array. ({\bf C}) Circuit diagram of the Kerr parametric oscillator, showing the SQUID array and shunting capacitor (black), coupling capacitor (blue) and transmission line (grey), and flux bias line (red).}
	\label{fig_device}
\end{figure}

The device, shown in Fig.~\ref{fig_device}A, was fabricated using a 5 mask lithography process on a 500-$\mu$m high-resistivity ($> 10~\text{k}\Omega \cdot \text{cm}$) Si substrate. First, the aluminum ground planes and feed lines are defined in photolithography using a liftoff process. Next, palladium marks are added in preparation for aligning subsequent electron-beam (e-beam) lithography masks. The SQUID array shown in Fig.~\ref{fig_device}B is fabricated using a Dolan-bridge double-angle technique for growing Al/AlO$_x$/Al junctions via {\it in situ} oxidation \cite{Dolan77,ss17}. After junction growth, the resonator capacitor is defined using e-beam lithography; narrow wires and an unconventional capacitor design were chosen to accommodate an array of nanomechanical resonators introduced in later devices \cite{arrangoiz18}, and are not essential to the design. Finally, a superconducting connection between the SQUID array and capacitor leads is formed using a bandage process \cite{Dunsworth2017}. An equivalent circuit diagram of the device is shown in Fig.~\ref{fig_device}C.

\subsection{Device parameters}
Table \ref{device_params} gives the device parameters for the Kerr parametric oscillator. Here, the maximum resonator frequency $\omegacmax$ is determined from a fit to the flux-bias tuning curve $\omegac(\Phi_e) = \omegacmax \sqrt{|\cos(\pi \Phi_e / \Phi_0)|}$,  where $\Phi_e$ is the externally applied magnetic flux and $\Phi_0$ is the flux quantum. The maximum Josephson energy of each of the $N=10$ SQUIDs, denoted by $\EJmax$ is determined from normal-state resistance measurements. Together, $\EJmax$ and $\omegacmax$ are used to extract the resonator charging energy $\EC$, which closely matches predictions from finite-element capacitance simulations. The resonator intrinsic and extrinsic decay rates are denoted as $\kappa_{\text{i}}$ and $\kappa_{\text{e}}$, respectively. $\chi$ is the resonator frequency shift per photon, determined from the peak-to-peak splitting in transient PSD measurements of a coherent state. The maximum parametric drive amplitude $\beta/2\pi$ used in the experiment is about 20 MHz. Larger $\beta$ are possible but lead to large states that cannot be faithfully reconstructed by the employed tomography procedure due to the nonlinearity of the system.
\begin{table}
	\centering
	\begin{tabular}{c c} \hline
		Parameter & Value \\ \hline
		$N$ & 10 \\
		$\omegacmax/2\pi$ & 8.35 GHz \\
		$\EC/h$ & 1.053 GHz \\
		$\EJmax/h$ & 82.79 GHz \\
		$\kappa_{\text{e}}/2\pi$ & 200 kHz \\
		$\kappa_{\text{e}}/2\pi$ & 900 kHz \\
		$\chi/2\pi$ & 17.3 MHz \\ \hline
	\end{tabular}
	\caption{Device parameters for the Kerr parametric oscillator.}
	\label{device_params}
\end{table}

\subsection{Experimental setup}\label{sec:expsetup}
\subsubsection{Up-conversion board}
We use two separate channels of a Tektronix series 5200 arbitrary waveform generator (AWG) to synthesize the temporal profile of the pulses at an intermediate frequency around 4 GHz and then further up-convert them to the desired frequencies close to the first (for the displacement) and second (for the parametric drive) harmonic of the resonator frequency using single sideband mixers (Fig.~\ref{fig_setup}). Both pulses are then combined together and sent to the sample through the flux line. Thanks to a weak but nonzero direct coupling of the flux line to the resonator, it can effectively double as a weakly coupled charge line which we can use instead of  sending the displacement pulses through the resonator input/output line. This way, we avoid saturation of the measurement setup by the reflection of the strong resonant pulse. The input/output line is not used for driving the system, except in initial characterization measurements of the resonator frequency using a vector network analyzer.

\subsubsection{Phase locking}\label{sec:phaselocking}
\begin{figure*}
	\centering
	\includegraphics[scale=0.7]{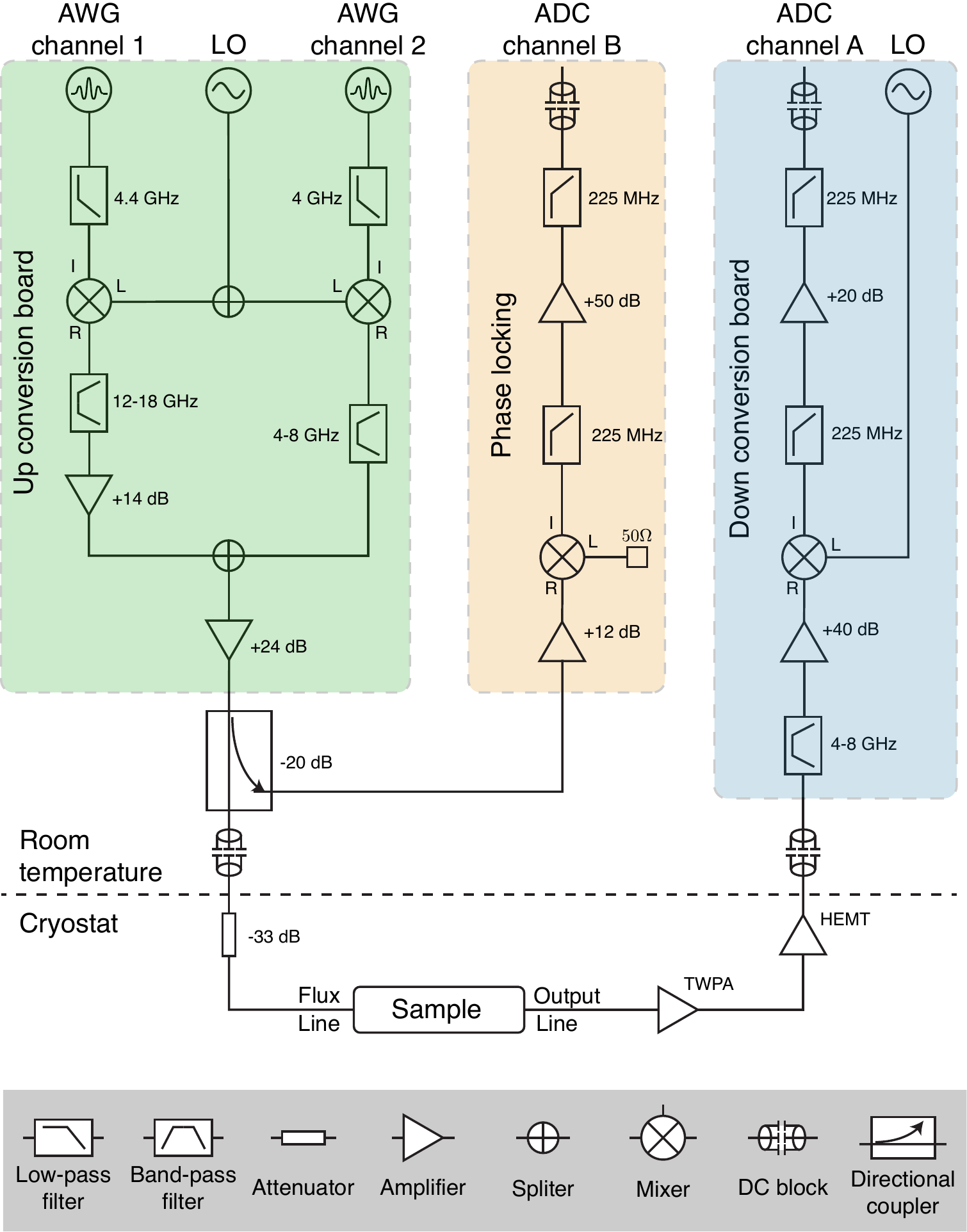}
	\caption{{\bf Diagram of the experimental setup.} }
	\label{fig_setup}
\end{figure*}

For a state prepared by parametric driving, the orientation of the reconstructed quasiprobability distribution in phase space is determined by the difference between the phase $\phid$ of the tomography pulses and the phase $\phip/2$ of the $\omegap/2$ subharmonic of the parametric driving. These in turn depend on the absolute phases of the up-conversion local oscillators. Slow changes of this relative phase due to phase drifts of the signal generators would lead to gradually accumulating errors over long measurements. To mitigate this, we monitor the phase over the course of the measurement using a second down-conversion board. Here we somewhat unconventionally feed both the parametric driving pulse at $\omegap$ and the displacement pulse at $\omegad$ into  the RF port of the mixer while the LO port is 50 Ohm terminated (Fig.~\ref{fig_setup}), relying on intermodulation to produce a signal at $\omegap-2\omegad$. By measuring this signal, we can determine the phase difference $\phi = \phip-2\phid$ which needs to be constant to ensure correct performance of the tomography measurements even with very long acquisition times. We observe a slow drift of $\phi$ of about 1 radian per hour which we then eliminate by measuring $\phi$ in approximately 1 minute intervals and appropriately adjusting the phase of the up-conversion LO.

\subsubsection{Down-conversion and digitization of the signal}
The signal emitted by the resonator at about 7 GHz is first amplified by a traveling wave parametric amplifier (TWPA) \cite{Macklin2015} and a low-noise high electron mobility transistor (HEMT) amplifier inside the cryostat. At room temperature, it is further amplified and then converted to an intermediate frequency of 125 MHz by a single-sideband mixer (Fig.~\ref{fig_setup}). This is then recorded by a digitizer card (AlazarTech ATS9350) with a 12-bit resolution and a sampling rate of 500 MS/s. The acquired data is first saved in the on-board memory buffer and then transferred to GPU for real-time data processing.

\section{Supplementary text}
\subsection{Hamiltonian of a parametric oscillator}
The Hamiltonian of the SQUID array resonator including the parametric driving is
\begin{equation}
\hat{H}/\hbar = 4 \EC \hat{n}^2 - N \EJ (\Phi(t)) \cos{\frac{\hat{\phi}}{N}}
\end{equation}
where $\hat{n}$ is the number of Cooper pairs and $\hat{\phi}$ the overall phase across the junction array. $\EC$ is the resonator's charging energy, $N$ is the number of SQUIDs in the array and $\EJ$ is the Josephson energy for a single SQUID in the array which depends on the external flux $\Phi(t)$. The flux is harmonically modulated around its mean value with a small amplitude such that $\EJ(\Phi(t))$ can be approximated as $\EJ + \delta\EJ \cos\omegap t$. After Taylor-expanding $\cos(\hat{\phi}/N)$ to fourth order, we get
\begin{equation}
\begin{split}
\hat{H}/\hbar = & 4 \EC \hat{n}^2 \\
& - N \EJ \left(
1 - \frac{1}{2} \left( \frac{\hat{\phi}}{N} \right)^2 + \frac{1}{24} \left( \frac{\hat{\phi}}{N} \right)^4 + \cdots
\right)\\
& - N\delta\EJ \left(
1 - \frac{1}{2} \left( \frac{\hat{\phi}}{N} \right)^2 + \cdots
\right) \cos\omegap t
\end{split}
\end{equation}

The quadratic time-independent part of the Hamiltonian can be diagonalized by defining
\begin{equation}
\begin{split}
\hat{n} &= -i n_0 (\hat{a}-\hat{a}^\dagger) \\
\hat{\phi} &= \phi_0 (\hat{a}+\hat{a}^\dagger).
\end{split}
\end{equation}
where $n_0^2 = \sqrt{\EJ/32N\EC}, \phi_0^2= \sqrt{2N\EC/\EJ}$ are the zero point fluctuations. We also drop c-valued terms in the expression above and get
\begin{equation}
\begin{split}
\hat{H}/\hbar =& \omegac^{(0)} \opad \opa - \frac{\EC}{12N^2} (\opa + \opad)^4 \\
&+ \frac{\delta\EJ \omegac^{(0)}}{4\EJ} 
(\opa+\opad)^2 \cos\omegap t,
\end{split}
\end{equation}
where $\omegac^{(0)} = \sqrt{8\EC\EJ/N} $.
We then transform the Hamiltonian into a rotating frame at the frequency $\omegap/2$, perform a rotating wave approximation assuming $|\omegac^{(0)} - \omegap/2| \ll \omegac^{(0)}$ and normal-order the resulting expression, which gives
\begin{equation}
\hat{H}/\hbar = \Delta \opad\opa - \frac{\chi}{2} \opad\opad\opa\opa + \beta (\opa^2 + \opa^{\dagger 2}),
\end{equation}
where we have defined the Kerr nonlinearity $\chi =\EC/N^2$, the parametric drive strength $\beta=\omegac^{(0)}\delta \EJ/8\EJ$, the dressed resonator frequency $\omegac=\omegac^{(0)}-\chi$ and the detuning $\Delta = \omegac - \omegap/2$.

\subsubsection{Effective potential}
\begin{figure}
	\centering
	\includegraphics[scale=0.6]{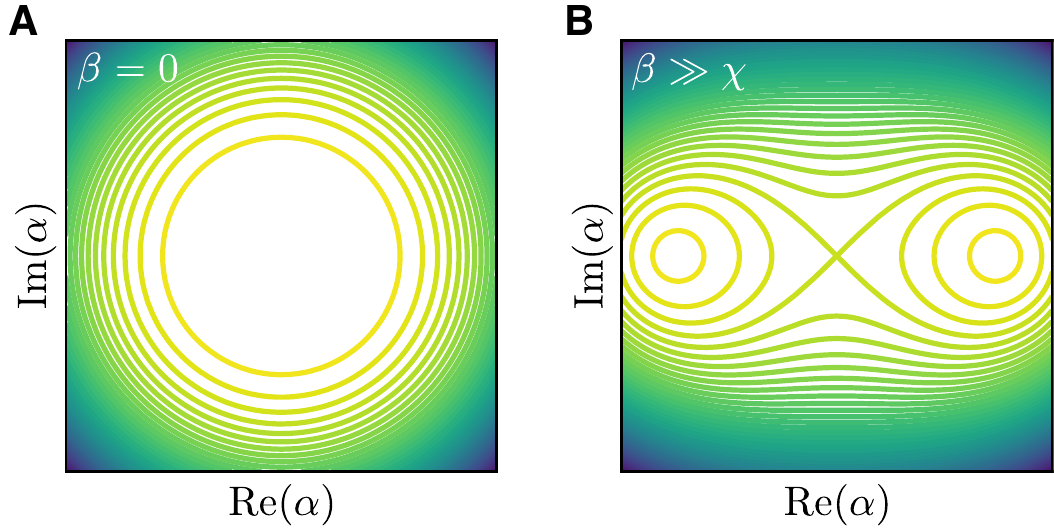}
	\caption{{\bf Illustration of the effective potential $V(\alpha)$ plot.} {\bf (A)} shows the potential at $\beta=0$ (below threshold) where there is only one global maximum while (B) represents the regime $\beta \gg \chi$ (far above threshold) where the potential has two symmetrically placed local maxima.}
	\label{fig_potential}
\end{figure}

An intuitive way to understand some aspects of the dynamics of this system is to define the following effective potential \cite{Zhang2017} in phase space
\begin{equation}
\begin{split}
V(\alpha) &= \mele{\alpha}{\hat{H}}{\alpha} = \Delta |\alpha|^2 - \frac{\chi}{2} |\alpha|^4 + \beta (\alpha^2 + \alpha^{* 2}) \\
&= |\alpha|^2 \left( \Delta - \frac{\chi}{2} |\alpha|^2 + 2\beta \cos 2\theta \right)
\end{split}
\end{equation}
where $\alpha = |\alpha | e^{i\theta}$. In the regime $2\beta > |\Delta|$, the effective potential has two symmetric local maxima (corresponding to classical stationary points of the system) at $|\alpha|>0$ and $\theta \in \{0,\pi\}$. Therefore in this case the effective potential has the shape of an inverted double-well. (Fig.~\ref{fig_potential}) When its maxima $\pm\alpha$ are sufficiently well separated, i.e., $\beta$ is large, the potential can be approximated by a quadratic function near $\pm\alpha$. Therefore the eigenstates of the system are close to Fock states displaced by $\pm\alpha$, as described in more detail below.

\subsubsection{Eigenstates of the Hamiltonian}
\begin{figure}
	\centering
	\includegraphics[scale=0.85]{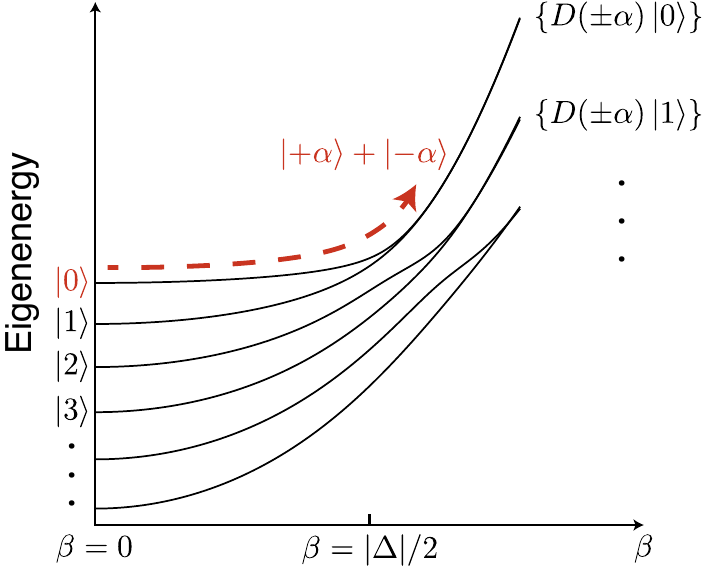}
	\caption{{\bf Eigenenergy of the Kerr parametric oscillator with different driving strength.} The plot is made in the classical limit $\chi \ll \Delta$ where the oscillation threshold happens at $\beta = | \Delta |/2$. The nearby energy levels merges together and becomes nearly degenerate at large $\beta$ where the degenerate subspaces are spanned by displaced Fock states. When we adiabatically increase $\beta$ from 0 to above threshold, the vacuum state will evolve into an even cat state since parametric driving preserves parity.}
	\label{fig_eigenenergy}
\end{figure}

Understanding the structure of the eigenvalues and eigenstates of the Hamiltonian at different $\beta$ helps us see how adiabatic ramping up of the drive amplitude can lead to Schr\"odinger's cat state generation \cite{Zhang2017}. At $\beta=0$, the eigenstates of the Hamiltonian are simply all Fock states $\{ \ket{n} \}$. As we increase $\beta$, two adjacent energy levels get closer and eventually merge together at very large $\beta$ (Fig.~\ref{fig_eigenenergy}). In the limit where $\beta/\chi \gg 1$, as suggested by the effective potential argument above, the eigenstates of the Hamiltonian form many two-dimensional nearly degenerate subspaces spanned by $\{ \hat{D}(\pm \alpha)\ket{n} \}$ for each $n$, where $\pm\alpha=\pm\sqrt{2\beta / \chi}$ are the locations of the local maxima. To show this, we first note that the coupling between states of the form $\hat{D}(+\alpha)\ket{n}$ and $\hat{D}(-\alpha)\ket{m}$ under the Hamiltonian $\hat{H}$ decreases exponentially with $|\alpha|^2$, i.e. 
\begin{equation}
\mele{m}{\hat{D}^\dagger (-\alpha)\hat{H}\hat{D}(\alpha)}{n} \sim e^{-2\alpha^2} \quad \forall m,n.
\end{equation}
Intuitively, this follows from the large separation of the two potential wells. The Hilbert space therefore effectively decomposes into two nearly decoupled subspaces consisting of states localized around $+\alpha$ and $-\alpha$. Next, we observe that the Hamiltonian within each of these subspaces is close to a harmonic oscillator. That is, $\hat{D}^\dagger (\alpha)\hat{H}\hat{D}(\alpha) \approx \hat{D}^\dagger (-\alpha)\hat{H}\hat{D}(-\alpha) \approx -2\chi \alpha^2 \opad \opa $ in the limit of $\alpha \gg 1$. Here we have used the relation $\alpha\approx\sqrt{2\beta / \chi}$ and only kept the highest order terms in $\alpha$. This confirms that displaced Fock states indeed approximate the pairwise degenerate eigenstates of $\hat{H}$ in the large $\alpha$ limit. 

It follows from the adiabatic theorem that if we prepare the system in the vacuum state and adiabatically increase $\beta$, the system will follow the eigenstate and end up within the corresponding degenerate subspace, as determined by the order of eigenenergies of the Fock states at $\beta=0$, which in turn depends on $\Delta$ and $\chi$. For $\Delta < 0$, $\ket{0}$ has the highest eigenenergy at $\beta=0$ and evolves into $\{ \hat{D}(\pm \alpha)\ket{0} \}$ at large $\beta$. Since parametric driving preserves the parity of the state, the final state is an even cat state $\ket{\alpha}+\ket{-\alpha}$. The case $\Delta > 0$ is more complicated since $\ket{0}$ may not have the highest eigenenergy at $\beta=0$ and can thus evolve into some $ [\hat{D}(\alpha) +\hat{D}(-\alpha)] \ket{n} $ where $n\neq 0$. Therefore for cat state generation, having a negative detuning is helpful since that gives a lager energy gap between $\ket{0}$ and all other higher Fock states and therefore allows a faster adiabatic tuning of $\beta$. Driving with an appropriately chosen positive detuning or non-adiabatic drive variations could on the other hand be useful for generating displaced Fock states or their superpositions \cite{Zhang2017}.


\subsection{Displacement pulse calibration}
The displacement pulses used in the experiments are 1 ns short pulses created by an arbitrary waveform generator. To calibrate them, we apply pulses with different amplitudes $\{V_i\}$ to the vacuum state of the resonator and measure the transient PSDs for the generated states. In order to simplify the analysis of the PSD measurements, we preprocess the raw data $\{\tilde{S}(\omega; V_i)\}$ by integrating over frequency bins centered around each of the individual transition peaks, thus effectively reducing the dimensionality of the analyzed data to $n\times m$, where $n$ is the number of bins and $m$ the number of different pulse amplitudes. 
\begin{equation}
\tilde{S}_j(V_i) = \int_{\text{bin }j}\tilde{S}(\omega;V_i)\,\mathrm{d}\omega
\end{equation}

The number of bins which can be usefully analyzed is limited by the increasing overlaps between the peaks corresponding to higher transitions with larger linewidths. When multiple transitions contribute to the same bin, the assumptions we use in our model to arrive at Eq.~(\ref{eq:binpower}) fail and a more complex model needs to be used. In most of our measurements, we have used $n = 4\text{ to }5$ bins.

\begin{figure*}
	\centering
	\includegraphics[scale=1.0]{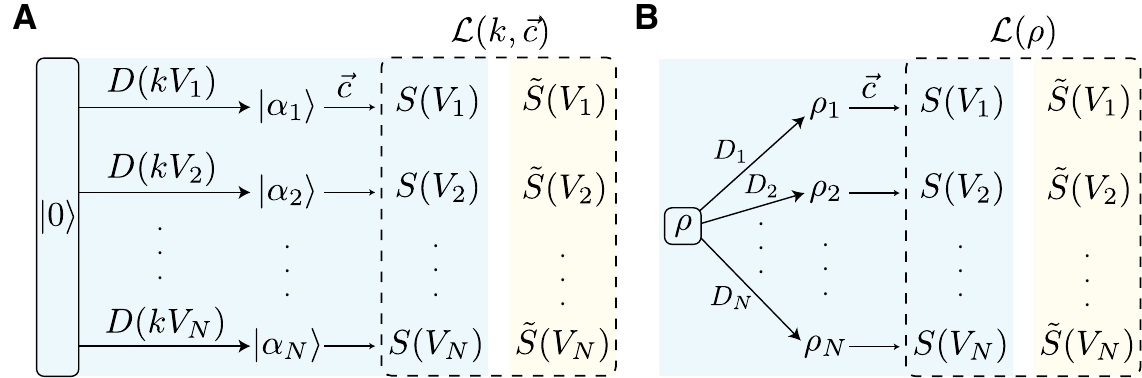}
	\caption{ {\bf Schematic of pulse calibration and tomography.} ({\bf A}) Pulse calibration is done by minimizing the loss function $\mathcal{L}(k, \vec{c})$, where $k$ is the scaling factor that converts voltage of the pulse into amplitude of displacement in phase space and $\vec{c}$ are the coefficients for the linear transformation that maps a quantum state into its transient PSD. ({\bf B}) The loss function for state tomography contains the contributions from the distance between measured and predicted transient PSDs for an unknown state $\hat{\rho}$ after each displacements.}
	\label{fig_pulse_cal}
\end{figure*}

The calibration is done under the following assumptions:
\begin{itemize}
	\item The state generated by a single pulse with voltage $V_i$ is a coherent state $\ket{\alpha_i}$.
	\item $\alpha_i$ depends linearly on the voltage $V_i$, i.e., $\alpha_i=k V_i$, where $k$ is a single fit parameter common to all pulses.
	\item The bin powers $\{\tilde{S}_j(V_i)\}$ calculated from the measured PSDs are related to the theoretical predictions $\{S_j(\ket{\alpha_i})\}$ by a gain factor $c_j$ which may in principle be different for each bin $j$.
\end{itemize}
The calibration parameters $k$ and $\vec{c} = (c_1,\ldots,c_n)$ are obtained by minimizing the loss function (Fig.~\ref{fig_pulse_cal}A)
\begin{equation}
\mathcal{L}(k, \vec{c}) = \sum_{i=1}^{m} \sum_{j=1}^{n} \left \Vert \tilde{S}_j(V_i) - c_j S_j(\ket{k V_i}) \right \Vert^2 .
\end{equation}
For a given $k$, finding optimal $\vec{c}$ reduces to a simple linear fitting problem and the value of $k$ is then calculated by minimizing $\mathcal{L}(k) = \min_{\vec{c}}\mathcal{L}(k, \vec{c})$.

To evaluate the loss function above, we need to calculate the theoretically expected total power $S_j(\hat{\rho}_0)$ in each transient PSD peak for a given state $\hat{\rho}_0$. The result represented by Eq.~(\ref{eq:binpower}) was outlined in the main text and its full derivation follows below.

\subsubsection{Analytical formula for the transient PSD}
The transient PSD for an initial state $\hat{\rho}_0$ is given formally by
\begin{widetext}
	\begin{equation}
	\begin{split}
	S(\omega; \hat{\rho}_0) = \int_0^{\infty} \text{d}t \int_0^{\infty} \text{d}t' \,\, \kappa \ave{\opad (t') \opa(t)} e^{-i\omega (t'-t)} = 2 \text{Re} \left \{ \int_{0}^{\infty} \text{d} t \int_{0}^{\infty} \text{d} \tau \,\, \kappa \ave{\opad (t+\tau) \opa(t)} e^{-i\omega\tau} \right \}
	\end{split}
	\end{equation}
\end{widetext}
which is normalized so that
\begin{equation}
\frac{1}{2\pi} \int_{-\infty}^{\infty} \text{d}\omega \, S(\omega; \hat{\rho}_0) = \ave{\opad (0) \opa(0)} .
\end{equation}
From the quantum regression theorem, we can rewrite the two-time correlation function as
\begin{equation}
\ave{\opad(t+\tau) \opa(t)} = \text{Tr} \left\{ \opad e^{\hat{\mathcal{L}}\tau} \opa e^{\hat{\mathcal{L}}t} \hat{\rho}_0 \right\}
\end{equation}
where $ \hat{\mathcal{L}} $ is the Liouvillian for the nonlinear resonator with the Hamiltonian  
\begin{equation}
\hat{H} = -\frac{\chi}{2} \opad\opad\opa\opa    
\end{equation}
and energy decay rate $\kappa$.

By solving the master equation and Fourier transforming the two-time correlation function, the analytical result for the transient PSD is
\begin{widetext}
	
	\begin{equation}
	S(\omega; \hat{\rho}_0) = 2 \text{Re} \left \{ \sum_{n=0}^{\infty} p_{n} \sum_{j=0}^{n-1} \frac{1}{(1+u)^{n-j-1}} \right. 
	\left.  \sum_{k=0}^{n-j-1} \left (
	\begin{array}{c}
	n-j-1 \\ k
	\end{array}
	\right)
	\frac{u^k}{i(\omega + k\chi) + 2(k+1)\kappa/2}
	\right \}
	\end{equation}
	
\end{widetext}
where $u = i\chi/\kappa$ and $p_n = \mele{n}{\hat{\rho}_0}{n}$. Notably, $S(\omega; \hat{\rho}_0)$ depends only on the diagonal part of $\hat{\rho}_0$ and this dependence is linear.

Our system satisfies $|u|=\chi/\kappa \gg 1$. In this regime, the transient PSD takes the form of a sum of relatively well separated peaks. The last equation can be approximated as
\begin{equation}
S(\omega; \hat{\rho}_0) \approx \sum_{n=0}^{\infty} p_{n} \sum_{j=0}^{n-1} 
\frac{(2j+1)\kappa}{(\omega + j\chi)^2 + [(2j+1)\kappa/2]^2}
\end{equation}
where the transient PSD is decomposed into a sum of Lorentzians with different linewidths and different center frequencies corresponding to the different peaks we measured. Therefore the total power in the $j$th peak is expected to be
\begin{equation}\label{eq:binpower}
S_j(\hat{{\rho}}_0) = \sum_{n=j}^{\infty} \mele{n}{\hat{\rho}_0}{n} = 1 - \sum_{n=0}^{j-1} \mele{n}{\hat{\rho}_0}{n}.
\end{equation}

\subsection{State tomography}
To reconstruct an unknown quantum state $\hat{\rho}$, we displace it in phase space (Fig.~3A) using short calibrated pulses with complex voltages $\{V_i\}$ (including phase), then measure the transient PSDs and calculate the corresponding integrated powers $\tilde{S}_j(V_i)$ for each of the bins $j=1,\ldots,n$ after each displacement $i=1,\ldots,m$. The density matrix $\hat{\rho}$ can be estimated by minimizing the difference between the predicted transient PSDs and the measured ones, which is expressed by the loss function (Fig.~\ref{fig_pulse_cal}B)
\begin{equation}
\mathcal{L}(\hat{\rho}) = \sum_{i} \sum_{j=1}^{n} \left \Vert \tilde{S}_j(V_i) - c_j S_j( \hat{D}(k V_i) \hat{\rho} \hat{D}^\dagger(k V_i) ) \right \Vert
\end{equation}
under the linear constraint $\text{Tr}(\hat{\rho}) = 1$ and under the condition that $\hat{\rho}$ is positive semidefinite. Notice that both the displacement and the map from a density matrix to a transient PSD are linear transformations and therefore this minimization problem is convex and can be efficiently solved by the Matlab package CVX \cite{cvx,gb08}.

\subsubsection{Parity constraint}
The form of the master equation of the parametrically driven system implies that all states which can be generated from a vacuum state are mixtures of states with even and odd parity. To see this, we only need to observe that the Hamiltonian conserves parity and the collapse operator $\sqrt{\kappa}\opa$ flips the parity of a state.

\begin{figure}
	\centering
	\includegraphics[scale=0.6]{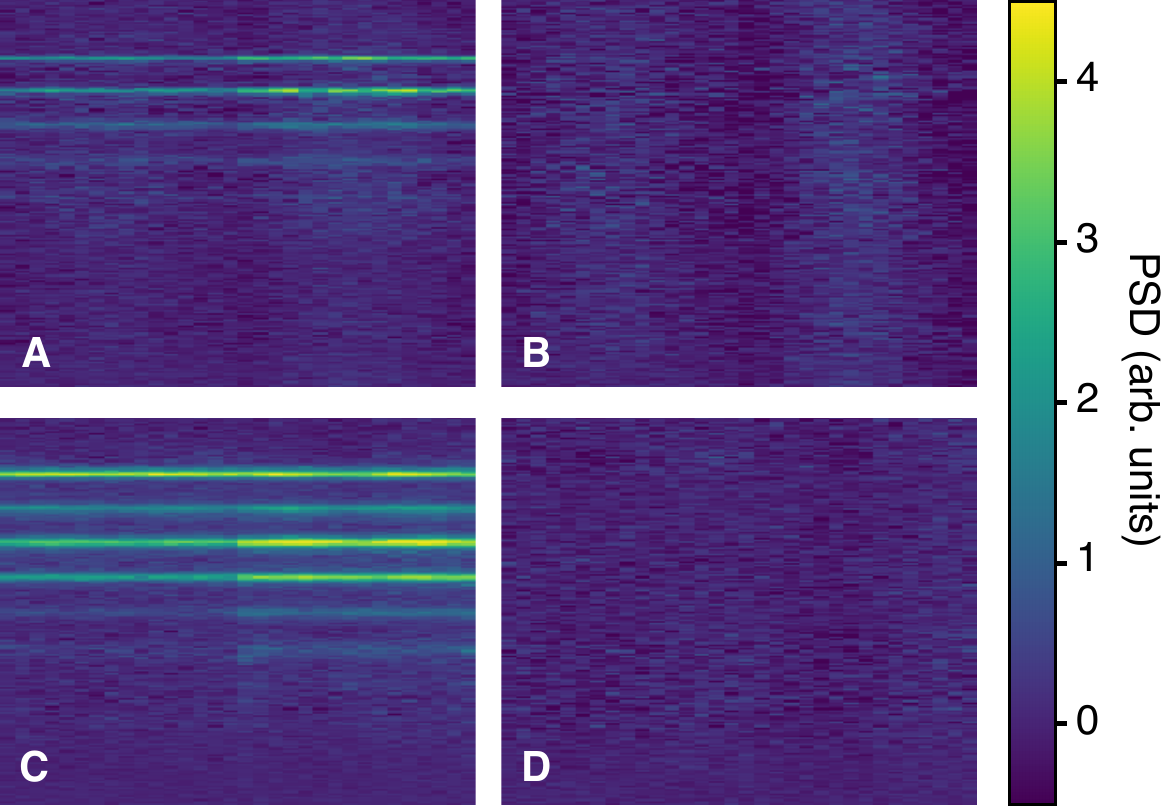}
	\caption{ {\bf Parity constraint.} ({\bf A}) Raw data $\{ \tilde{S}(\omega,\alpha) \}$ of (Fig.~4B left), including transient PSDs for all displacements $\alpha$. ({\bf B}) The difference of the raw data under parity transformation $\{\tilde{S}(\omega,+\alpha) - \tilde{S}(\omega,-\alpha) \}$. ({\bf C-D}) are the same as ({\bf A-B}) but for the tomography dataset in (Fig.~4B right).}
	\label{fig_parity}
\end{figure}

When reconstructing states prepared by parametric driving, we use this condition as an additional constraint on the unknown density matrix $\hat{\rho}$, requiring that
\begin{equation}\label{eq:parityconstraint}
\hat{P}\hat{\rho} \hat{P}^\dagger = \hat{\rho},
\end{equation}
where $\hat{P} = e^{i\pi \opad \opa}$ is the parity operator. We justify this assumption by verifying the corresponding symmetry of the transient PSDs measured in the tomography process under a rotation of the applied displacement by $\pi$. For a state $\hat{\rho}$ satisfying Eq.~(\ref{eq:parityconstraint}), we expect that $\tilde{S}(\omega;+\alpha) = \tilde{S}(\omega;-\alpha)$ and we check this by plotting the difference $\tilde{S}(\omega;+\alpha)-\tilde{S}(\omega;-\alpha)$ and observing that it is negligible when compared with the PSDs $\tilde{S}(\omega;\alpha)$ themselves (Fig.~\ref{fig_parity}).

\subsection{Data fitting}
Some parameters of the system such as its nonlinearity $\chi$ and linewidth $\kappa$ do not change significantly among different experiments and we therefore assume constant values for them which are obtained from initial characterization measurements. Other parameters like the detuning $\Delta$ and the parametric driving amplitude $\beta$ vary between experiments and therefore their values are determined separately in each instance, either directly from the settings of the experiment or by fitting.

The nonlinearity $\chi/2\pi=17.3\,\mathrm{MHz}$ was calculated from transient PSD measurements as the mean spacing between adjacent peaks. To fit the time evolution of the photon number (Fig.~4A), we introduce the detuning $\Delta$, the linewidth $\kappa$ and the drive conversion factor $\beta/V$ that relates the voltage amplitude $V$ applied to the flux line to the parametric driving amplitude $\beta$ as fit parameters and get $\Delta/2\pi=24.6\,\mathrm{MHz}$ and $\kappa/2\pi = 1.1\,\mathrm{MHz}$ by minimizing the $L^2$ distance between the simulation results and the measured data. The value of $\Delta/2\pi$ found by fitting is very close to the value $24\,\mathrm{MHz}$ set in the experiment and the small difference is likely due to slow magnetic flux noise which causes variations in the resonator frequency. The linewidth is also consistent with direct VNA measurements, which give values around $1\,\mathrm{MHz}$, slightly depending on the resonator frequency. Since none of the results in this work are very sensitive to the exact value of the linewidth, we fix $\kappa/2\pi$ to be $1.1\,\mathrm{MHz}$ in all subsequent theory fits.

The process of fitting the steady-state mean photon number $\overline{n}$ (Fig.~2C) is similar to the case of the time-dependent $\overline{n}$ measurement described above (Fig.~4A) except that $\kappa$ is fixed and the detuning resulting in the best fit is $\Delta/2\pi=25.3\,\mathrm{MHz}$. For fitting the PSDs at steady state (Fig.~2D), we fix $\Delta/2\pi$ to the value $11.2\,\mathrm{MHz}$ set in experiment. The only fit parameter is the drive conversion factor $\beta/V$.

The tomography measurement of the freely evolving state (Fig.~3C) has two fit parameters: the size and phase of the coherent state at $\tau=0$. In principle, the observed phase $\theta$ of the state should be easily predictable since it only depends on the relative phase between the preparation pulse and the tomography pulse, both of which are generated by the AWG and undergo the same path through the up-conversion chain. A calculation based on the used experimental settings gives a predicted phase of $\theta\approx 1.19\pi$. The size of the state $\alpha$ can also be estimated from the pulse calibration parameter $k$, which gives $\alpha \approx 1.5$. Treating both $\theta$ and $\alpha$ as unknown fit parameters, we get $\theta\approx 1.24\pi$ and $\alpha \approx 1.0$. In this fitting, to achieve simultaneous match to the different states at each of the different evolution times $\tau$, we choose the objective function to be the geometric mean of the fidelities between each measured state and the corresponding theoretical prediction.

For states prepared by parametric driving, their phase $\theta$ depends on the absolute phase of the signal generator. Through the feedback loop described in Section \ref{sec:phaselocking}, we stabilize the phase over the measurement time at a fixed value. This value in principle depends on the frequency of the signal in a complex way due to the variation of the system's S-parameters with frequency. Since the different tomography measurements are mostly performed at different resonator frequencies, we treat the phase of the state $\theta$ at each of these frequencies as a fit parameter. Another fit parameter we introduce for these measurements is the possible small time interval $\td$ of free evolution between the preparation pulse and the tomography pulses. This is to account for a potential delay between the two pulses which are generated by different channels of the AWG and processed by separate up-conversion boards.

For the tomography measurements of the steady state under parametric driving (Fig.~4B right), both the detuning $\Delta$ and the parametric driving amplitude $\beta$ have been fixed through fitting to the time evolution of the photon number (Fig.~4A). Therefore the only remaining fit parameters are the phase of the state and the delay time $\td$. By maximizing the fidelity between the reconstructed density matrix and the simulation result, we get $\td = 2.5\,\mathrm{ns}$, which is then kept fixed for all other tomography measurements with parametric driving. Consequently, for the transient state at $20\,\mathrm{ns}$ (Fig.~4B left), the only fit parameter is its phase $\theta$.

For the tomography measurements of the adiabatically prepared cat states, the phase $\theta$ is again unknown but should be the same for all four states. Therefore the phase $\theta$ and the drive conversion factor $\beta/V$ are the only fit parameters for all four data sets. We again assume that the final state is distorted by a short period of free evolution whose length we fix at $\td = 2.5\,\mathrm{ns}$ based on previous measurements.




\end{document}